\documentclass[10pt,superscriptaddress,tightenlines,twocolumn,amsmath,amssymb,aps, pra]{revtex4-2}
\UseRawInputEncoding
\usepackage{graphicx}
\usepackage{dcolumn}
\usepackage{bm}
\usepackage{color}
\usepackage{amsmath}
\usepackage{amssymb}
\usepackage{subfigure}
\usepackage{braket}
\usepackage{indentfirst}
\usepackage{mathrsfs}
\usepackage[linesnumbered,ruled,vlined]{algorithm2e}
\usepackage{enumerate}
\usepackage{enumitem} % for item without label
\usepackage{}
\usepackage{algorithmic}
\usepackage{amsfonts}
\usepackage{hyperref}
\usepackage{makecell}
\usepackage{lmodern}
\usepackage{lineno}
\usepackage{multirow}

\newcommand{\tn}[1]{\textnormal{#1}}
\newcommand{\be}{\begin{equation}}
\newcommand{\ee}{\end{equation}}

\newcommand{\esec}{\eps_{\tn{sec}}}

\newcommand{\sketbra}[2]{{\ensuremath{\lvert #1\rangle\!\langle #2\rvert}}}
\newcommand{\lketbra}[2]{{\ensuremath{\left\lvert #1\right\rangle\!\!\left\langle #2\right\rvert}}}
\newcommand{\ketbra}[2]{\if@display\lketbra{#1}{#2}\else\sketbra{#1}{#2}\fi}

\newcommand{\one}{\ket{1}}

\newcommand{\eps}{\varepsilon}

\newcommand{\cK}{\mathcal{K}}

\newcommand{\rs}{{\rm{1}}}
\newcommand{\rd}{{\rm{2}}}
\newcommand{\rdd}{{\rm{3}}}

\makeatletter

\newcommand{\Rmnum}[1]{\expandafter\@slowromancap\romannumeral #1@}
\makeatother

\usepackage{xcolor}
\colorlet{RED}{red}
\colorlet{BLACK}{black}

\providecommand{\U}[1]{\protect\rule{.1in}{.1in}}
\hypersetup{colorlinks=true, linkcolor=red, citecolor=blue, urlcolor=black}
\begin{document}

\title{Information-theoretically secure quantum timestamping with\\one-time universal hashing}
\author{Ming-Yang Li}\thanks{These authors contributed equally to this work.}
\affiliation{National Laboratory of Solid State Microstructures and School of Physics, Collaborative Innovation Center of Advanced Microstructures, Nanjing University, Nanjing 210093, China}

\author{Chen-Xun Weng}\thanks{These authors contributed equally to this work.}
\affiliation{National Laboratory of Solid State Microstructures and School of Physics, Collaborative Innovation Center of Advanced Microstructures, Nanjing University, Nanjing 210093, China}

\author{Wen-Bo Liu}\thanks{These authors contributed equally to this work.}
\affiliation{College of Information and Communication, National University of Defense Technology, Wuhan 430010, China}

\author{Mengya Zhu}
\affiliation{MatricTime Digital Technology Co. Ltd., Nanjing 211899, China}

\author{Zeng-Bing Chen}\email{zbchen@nju.edu.cn}
\affiliation{National Laboratory of Solid State Microstructures and School of Physics, Collaborative Innovation Center of Advanced Microstructures, Nanjing University, Nanjing 210093, China}
\affiliation{MatricTime Digital Technology Co. Ltd., Nanjing 211899, China}

\date{\today}
\begin{abstract}
Accurate and tamper-resistant timestamps are essential for applications demanding verifiable chronological ordering, such as legal documentation and digital intellectual property protection. Classical timestamp protocols rely on computational assumptions for security, rendering them vulnerable to quantum attacks, which is a critical limitation given the rapid progress in quantum computing. To address this, we propose an information-theoretically secure quantum timestamping protocol based on one-time universal hashing with quantum keys. Our protocol simultaneously achieves information-theoretic security and high efficiency, enabling secure timestamping for arbitrarily long documents. Simulations demonstrate a generation rate exceeding 100 timestamps per second over intercity distances. In addition, our protocol only requires weak coherent states, making it practical for large-scale deployment. This work advances the field of quantum timestamping and contributes to the broader development of quantum cryptography and the future quantum internet.
\end{abstract}
 
\maketitle

\section{Introduction}\label{section1}

Accurate and reliable timestamps are crucial for determining the chronological order of events in various applications. They provide legally and technically precise time values, ensuring that issuance, query, and verification processes comply with the non-repudiation and unforgeability requirements of network applications. Secure and reliable timestamp protocols are crucial for legal documentation, digital intellectual property protection, and data integrity verification. For example, in electronic financial systems, such as electronic banking or Alipay, a precise completion time of the ownership transfer for a significant asset is legally crucial, which consists of reliable proof of when the buyer's ownership of the asset starts. To support these requirements, various classical timestamp protocols utilizing hash algorithms and digital signatures have been proposed and implemented in practice~\cite{Bayer1993Improving,Zhang2020Chronos,Estevam2021Accurate,wu2023distributed,Culot2019addressing}. However, these existing classical protocols face a fundamental security challenge: the vulnerability of classical digital signature schemes and computational assumptions to quantum attacks given the ongoing advancements in quantum computing and algorithms~\cite{shor1994algorithms,shor1999polynomial,grover1997quantum,martin2012experimental,fedorov2018quantum,Gouzien2021Factoring,zhou2022experimental}. This weakness creates opportunities for timestamp manipulation of malicious attackers during issuance and transmission, compromising the reliability of the received timestamp and undermining its credibility.

On the other hand, based on the basic principles of quantum mechanics, quantum cryptographic protocols offering information-theoretic security and superior performance have been developed, such as quantum key distribution (QKD)~\cite{bennett1984quantum,yin2016measurement,lucamarini2018overcoming,wang2018twin-field,xie2022breaking,Li2024finite,Zhong2024Hyperentanglement,zhan2025experimental}, quantum secure direct communication~\cite{pan2024Evolution,Sun2025Quantum,pan2025Simultaneous}, quantum digital signatures (QDS)~\cite{dunjko2014quantum,yin2016practical,qin2022quantum,lu2021efficient,Weng2021secure,yin2023experimental,li2023one}, quantum Byzantine agreement~\cite{Fitzi2001quantum,gaertner2008xperimental,kiktenko2018quantum,weng2023beating,jing2024experimental}, quantum e-commerce~\cite{cao2024Experimental} and quantum digital payments~\cite{pastawski2012unforgeable,guan2018Experimental,bozzio2018experimental,Bozzio2019Semi,horodecki2020Semi,schiansky2023demonstration}. However, to date, no existing information-theoretically secure quantum timestamp protocols can guarantee transferability, unforgeability, and non-repudiation of timestamps against quantum attackers without relying on computational assumptions. Developing such a quantum solution to protect timestamps against quantum attackers is crucial for secure timestamping in the quantum era.

Here, for the first time, we present a quantum timestamp protocol offering information-theoretic security against quantum attacks and high efficiency for documents of arbitrary length. By using the one-time universal hashing (OTUH) introduced in Ref.~\cite{yin2023experimental} (see Appendix~\ref{Sec:A1}), the protocol employs different quantum keys for each hash computation, making it infeasible for the hash values to be attacked. Our protocol employs quantum key generation (QKG) to produce correlated quantum secret keys for one-time universal hash function construction and one-time pad encryption, ensuring security against quantum attacks without relying on computational assumptions. Moreover, we prove the composable security of our protocol, demonstrating the resilience against both internal and external threats. \textcolor{black}{The security of the timestamps is ensured by the non-repudiation and unforgeability properties derived from the inherent asymmetric correlations of quantum keys in our protocol~\cite{yin2023experimental,li2023one}.} Numerical simulations demonstrate that our protocol can achieve a rate exceeding 100 secure timestamps per second over intercity distances. The protocol requires only the preparation of weak coherent quantum states, not entangled states or high-dimensional qudits, and is compatible with QKG stages adapted from various QKD protocols, simplifying its large-scale experimental deployment.

The remaining part of this paper is organized as follows. Section~\ref{sec:2} describes the scheme of our timestamp protocol. Section~\ref{sec:3} presents details of security analysis. Numerical simulations are presented in Sec.~\ref{sec:4} to show the efficiency of our scheme. Finally, we conclude in Sec.~\ref{sec:5}.

\section{Protocol description}\label{sec:2}

In our quantum timestamp scheme, there are \textcolor{black}{four legal parties: a subscriber, a verifier, a trusted certificate authority (CA), and a timestamp authority (TSA)}. \textcolor{black}{Our scheme considers the scenario where the subscriber possesses a document, $doc$, and seeks a timestamp to prove its existence prior to a specific point in time. The verifier checks the validity of the file and its timestamp. CA records the file to prevent unauthorized alteration by the subscriber. Finally, TSA provides the timestamp and its associated signature. Through our protocol, a reliable timestamp can be created for a valid file. After successful verification, the file and its timestamp can not be tampered with or denied.}

The whole protocol includes two stages. First, the related participants share correlated quantum keys through a QKG stage. Then, \textcolor{black}{a timestamp is generated for a certified file; the file's integrity and the timestamp's validity are verified using three-party verification with the messaging stage of the OTUH-QDS~\cite{yin2023experimental,li2023one}.}

\subsection{Quantum key generation stage}
\textcolor{black}{The QKG stage is employed to generate correlated quantum keys for three participants, and any QKD and quantum secret sharing~\cite{hillery1999quantum,cleve1999how,wei2013experimental,williams2019quantum,de2020experimental,gu2021differential,shen2023experimental,li2023breakingqss,Xiao2024Source,wang2024Experimental} protocol can be adapted for this QKG stage~\cite{yin2023experimental,li2023one}.} We illustrate this stage using the two-photon twin-field (TPTF)-QKG as an example, adapted from the two-party TPTF-QKD protocol in Ref.~\cite{xie2023Scalable}. Simulation results for sending-or-not-sending (SNS)-QKG (based on SNS-QKD in Ref.~\cite{jiang2019unconditional}) and BB84-QKG (based on BB84-QKD in Ref.~\cite{lim2014concise}) are also provided. Detailed calculations for these three QKG methods are given in Appendix~\ref{Sec:A4}. 

The steps for TPTF-QKG are detailed below. The three participants are named $U_1$, $U_2$ and $U_3$ respectively.

\textbf{(a) Preparation.} For each of the $N_{\text{tot}}$ time bins, $U_1$ and $U_2$ independently generate a weak coherent pulse. These pulses are encoded with randomly selected phases ($\theta_x^l \in [0, 2\pi)$) and bits ($r_x^l \in \{0, 1\}$), where $x \in \{1, 2\}$ indexes the users and $l \in \{1,2,\cdots,N_{\text{tot}}\}$ denotes the sequence of time bin. The pulses are described by the state $|\psi_x^l\rangle = |e^{i(\theta_x^l + r_x^l \pi)}\sqrt{k_x^l}\rangle$, with intensity $k_x^l$ drawn probabilistically ($p_{k_x}$) from the set $\{\mu_x, \nu_x, \mathbf{o}_x, \hat{\mathbf{o}}_x\}$. This set represents signal, decoy, and two types of vacuum states (preserved and declared), satisfying $\mu_x > \nu_x > \mathbf{o}_x = \hat{\mathbf{o}}_x = 0$. Following encoding, the pulses are transmitted to an untrusted relay (UR) over insecure channels. Concurrently, a bright reference pulse is also sent to the UR to enable measurement of the phase noise difference $\phi_{12}^l$. This establishes the initial state for the subsequent stages of the protocol.

\textbf{(b) Measurement.} For UR, interference measurements are performed on each arriving pulse pair. A beam splitter and two detectors are employed; a successful detection event is defined as only one detector registering a click. The UR publicly announces each successful detection event, identifying the specific detector that registered the click. For clarity, we use a notation that includes the intensity selections of both users. For example, $\{\mu_1, \nu_2\}$ indicates that user $U_1$ selected the signal intensity ($\mu_1$), and user $U_2$ chose a decoy intensity ($\nu_2$).

\textbf{(c) Sifting.} Successful detection events are categorized into two distinct sets. The first set comprises events where neither user employs decoy or declared-vacuum intensities; these are $\{\mu_1, \mathbf{o}_2\}$, $\{\mu_1, \mu_2\}$, $\{\mathbf{o}_1, \mu_2\}$, and $\{\mathbf{o}_1, \mathbf{o}_2\}$, and are used to generate the Z-basis key. The second set encompasses all remaining successful events and is dedicated to parameter estimation for the X-basis. For Z-basis key generation, $U_1$ randomly pairs a time bin ($l_1$) with intensity $\mu_1$ and another time bin ($l_2$) with intensity $\mathbf{o}_1$. A bit value of 0 (1) is assigned if $l_1 < l_2$ ($l_1 > l_2$), with $l_1$ and $l_2$ communicated to $U_2$. If $U_2$'s intensities are $k_2^{\min(l_1, l_2)} = \mu_2$ ($\mathbf{o}_2$) and $k_2^{\max(l_1, l_2)} = \mathbf{o}_2$ ($\mu_2$), $U_2$ assigns a bit value of 0 (1). Events with $k_2^{l_1} = k_2^{l_2} = \mathbf{o}_2$ or $\mu_2$ are discarded. The resulting Z-basis events are $\{\mu_1 \mathbf{o}_1, \mathbf{o}_2 \mu_2\}$, $\{\mu_1 \mathbf{o}_1, \mu_2 \mathbf{o}_2\}$, $\{\mathbf{o}_1 \mu_1, \mathbf{o}_2 \mu_2\}$, and $\{\mathbf{o}_1 \mu_1, \mu_2 \mathbf{o}_2\}$. X-basis parameter estimation involves exchanging intensity and phase information via an authenticated channel. The global phase difference at time bin $l$ is $\theta^l = \theta_1^l - \theta_2^l + \phi_{12}^l$. Events $\{\nu_1^l, \nu_2^l\}$ are retained only if $\theta^l \in [-\delta, \delta] \cup [\pi - \delta, \pi + \delta]$, where $\delta$ defines the phase slice. Pairs of retained events with $|\theta^{l_1} - \theta^{l_2}| = 0$ or $\pi$ are formed: $\{\nu_1^{l_1} \nu_1^{l_2}, \nu_2^{l_1} \nu_2^{l_2}\}$. $X$-basis bit values are derived from $r_1^{l_1} \oplus r_1^{l_2}$ and $r_2^{l_1} \oplus r_2^{l_2}$. $U_2$ always flips its Z-basis bit. In the X-basis, $U_2$ conditionally flips bits to ensure correlation with $U_1$: If the global phase difference is 0 ($\pi$) and the UR detectors are different (same), $U_2$ flips; otherwise, it keeps its bit. The remaining events are used for decoy-state analysis.

\textbf{(d) Parameter estimation.} From the randomly selected Z-basis bits, $U_1$ and $U_2$ generate an $n^Z$-bit raw key. The remaining Z-basis bits are used to estimate the bit error rate, $E^Z$. Following this, all X-basis bit values are exchanged to count the total number of errors. Employing a decoy-state method~\cite{wang2005beating,lo2005decoy}, the number of vacuum events ($s_{0\mu_2}^Z$), single-photon pairs ($s_{11}^Z$), and the single-photon pair phase error rate ($\phi_{11}^Z$) in the Z-basis are estimated.

\textbf{(e) Post-processing.} The key post-processing involves two steps.
\begin{enumerate}
    \item \textit{Error Correction and Verification}: An error correction algorithm with $\varepsilon_{\text{cor}}$-correctness ~\cite {Brassard1994Secret,yan2008information} is applied by $U_1$ and $U_2$ to distill the final key. The resulting key remains $n^Z$ bits long. To further enhance security, $U_1$ permutes the key and transmits this permutation to $U_2$ securely. 
    \item \textit{Privacy Amplification}: A randomly chosen universal$_2$ hash function is applied by $U_1$ to the $n^Z$-bit error-corrected key, producing an $n^{Z'}$-bit final key. This function is then communicated to $U_2$, who applies the same function to their $n^Z$-bit key to obtain the final $n^{Z'}$-bit key. This step significantly reduces the information potentially accessible to an eavesdropper.
\end{enumerate}

\textcolor{black}{\textbf{(f) Correlating.} $U_1$ and $U_3$ conduct steps \textbf{(a)}--\textbf{(e)} again to generate another quantum key with equal length. Then, $U_1$ performs a bitwise XOR operation on the two resulting keys.}

For the timestamp stage, the following key generations are required: \textcolor{black}{(1) The subscriber and verifier share a $3n$-bit key, divided into an $n$-bit segment ($s_1$) and a $2n$-bit segment ($r_1$). The subscriber and CA also share a $3n$-bit key, partitioned into an $n$-bit segment ($s_2$) and a $2n$-bit segment ($r_2$). The subscriber then calculates $s = s_1 \oplus s_2$ and $r = r_1 \oplus r_2$ to form related quantum keys, where $\oplus$ denotes the bit-wise XOR of two strings.} \textcolor{black}{(2)} TSA and subscriber share a $3m$-bit key, divided into an $m$-bit segment ($u_1$) and a $2m$-bit segment ($v_1$). TSA and verifier also share a $3m$-bit key, partitioned into an $m$-bit segment ($u_2$) and a $2m$-bit segment ($v_2$). TSA then calculate $u = u_1 \oplus u_2$ and $v = v_1 \oplus v_2$ to form related quantum keys.

 \begin{figure*}[t]
    \centering
    \includegraphics[width=\textwidth]{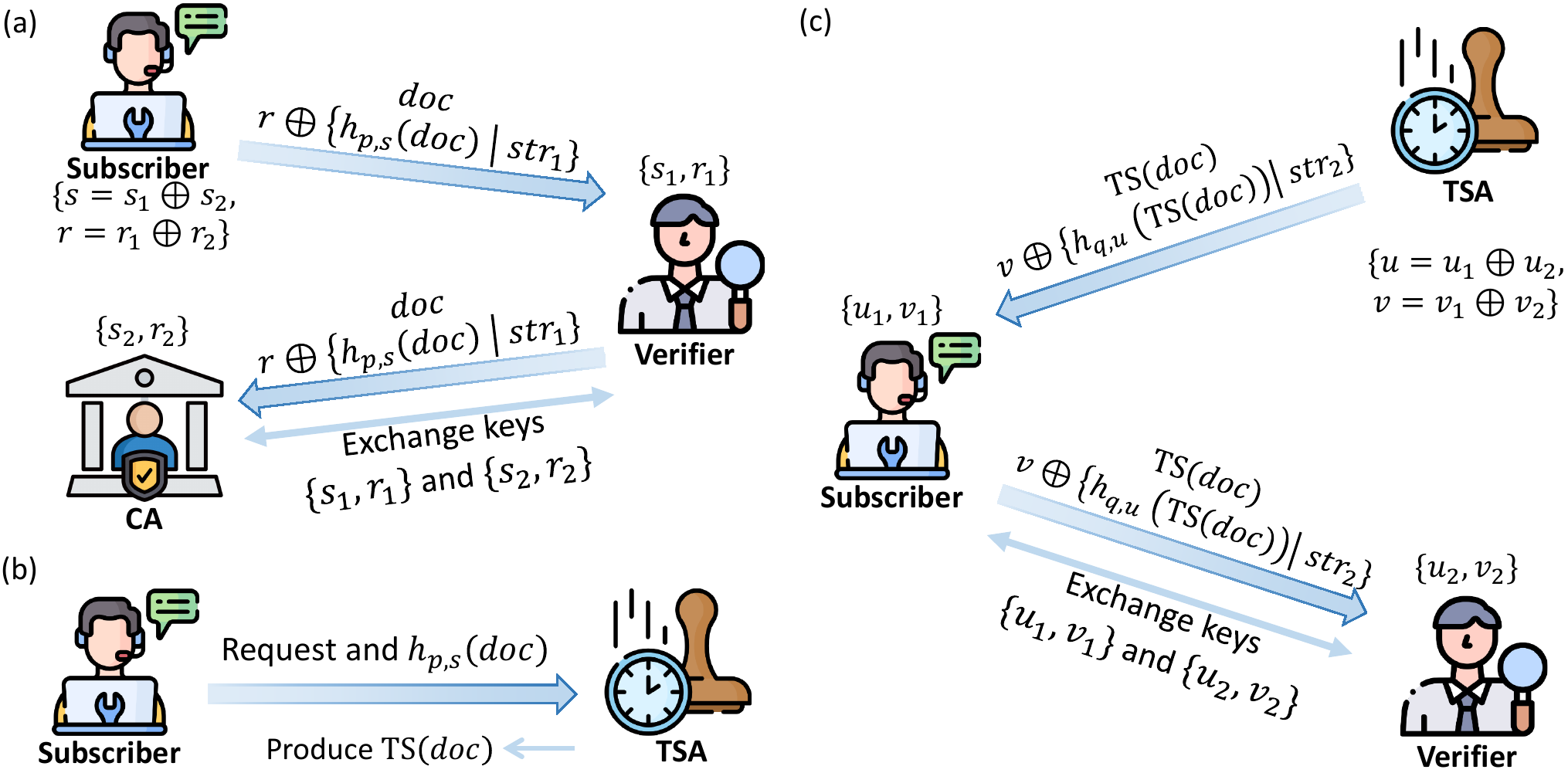}
    \caption{Illustration of our quantum timestamp scheme. (a) File verification: The subscriber generates a hash digest of the document, encrypts it along with the irreducible polynomial coefficients, and sends it to the verifier for authentication. \textcolor{black}{The verifier forwards the document and its corresponding encrypted hash value to CA and exchanges quantum keys with CA; both of them independently verify the validity of the file and its corresponding hash value according to the exchanged quantum keys.} (b) Timestamp generation: Upon successful verification, TSA generates a timestamp document incorporating the verified digest, timestamp, request ID, and the identities of the authority and verifier. (c) Timestamp verification: TSA sends the timestamp and its encrypted hash to the subscriber, who forwards it to the verifier and exchanges keys; both independently verify the hash, ensuring timestamp validity.} %\textcolor{black}{Classical authenticated channels are required for communication between the verifier and CA during file verification, and between the subscriber and verifier during timestamp verification.}
    \label{fig1}
\end{figure*}

\subsection{Timestamp stage}

The timestamp stage includes three steps. In the first step, \textcolor{black}{The subscriber, the verifier and CA check the correctness of the digest of $doc$ with CA recording the file.} In the second step, TSA produces the timestamp file according to the digest and the time. \textcolor{black}{TSA, the subscriber, and the verifier} then verify the validity of the timestamp in the third step. The details of our proposed timestamp stage are presented below and illustrated in Figure~\ref{fig1}. The length of the file $doc$ is denoted as $L$.

\textbf{Step 1. File verification.}

(a) The subscriber generates a binary random bit string $str_1$ with length $n$ with the local quantum random number generator (QRNG)~\cite{lunghi2015selftesting,cao2016source,miguel2017quantum,liu2018device,liu2023source}. $str_1$ is used to generate an irreducible polynomial $p(x)$ of order $n$. The specific method of generating $p(x)$ is provided in Appendix~\ref{Sec:A2}.

(b) The subscriber generates the LFSR-based Toeplitz hash function $h_{p,s}$ by $p(x)$ and $s$ as input random number, as shown in Appendix~\ref{Sec:A3}. Then the hash function $h_{p,s}$ is used to hash the file $doc$ to be timestamped, and the digest $h_{p,s}(doc)$ is obtained. The subscriber encrypts $h_{p,s}(doc)$ and $str_1$ with the shared key $r$ to obtain $r \oplus \{ h_{p,s}(doc)~|~str_1 \}$ and sends it to the verifier. The file $doc$ is also sent to the verifier without encryption. \textcolor{black}{The verifier then forwards $doc$ and $r \oplus \{ h_{p,s}(doc)~|~str_1 \}$ to CA.}

(c) \textcolor{black}{The verifier and CA exchange their quantum keys $s_1,s_2,r_1,r_2$ with each other via classical authenticated channels. Each of them can combine $s_1$ and $s_2$ to get $s^{\prime}$, $r_1$ and $r_2$ to get $r^{\prime}$. Subsequently, the verifier and the CA utilize $r^{\prime}$ to decrypt the ciphertext, thereby retrieving $h_{p, s} \left ( doc \right )$ and $str_1$. The polynomial $p^{\prime}(x)$ is generated using $str_1$ and a hash function $h_{p, s}^{\prime}$ is obtained using $p^{\prime}(x)$ and $s^{\prime}$. The verifier and CA apply $h_{p, s}^{\prime}$ to $doc$ and verify the equivalence of the hash value $h_{p, s}^{\prime}\left (doc\right )$ with $h_{p, s} \left ( doc \right )$. If they are the same, the authentication succeeds. If they are different, the authentication fails.}

\textcolor{black}{(d) The file is deemed valid only after successful verification by both parties.}

\textbf{Step 2. Timestamp production.}

(a) After successful authentication, the subscriber sends a timestamp request to TSA and attaches the digest $h_{p,s}(doc)$ verified by the verifier.

(b) TSA gets the time value $T$ at the requested time from the National Time Centre.

(c) Given that there may be more than one subscriber request timestamp within a standard time precision, multiple requests are numbered by a Request identifier: $\operatorname{Req-ID}$. The assignment of $\operatorname{Req-ID}$ is random, not sequential.

(d) With the organization identifier $\operatorname{Auth-ID}$ of TSA and organization identifier $\operatorname{Ver-ID}$ of the verifier, TSA produces the timestamp document $\operatorname{TS}(doc)$ :

\begin{equation}
    \operatorname{TS}(doc)=\left[ h_{p, s}(doc), T, \operatorname{Req-ID}, \operatorname{Auth-ID}, \operatorname{Ver-ID} \right ],
\end{equation}
whose length is denoted as $k$.

The inclusion of $\operatorname{Auth-ID}$ and $\operatorname{Ver-ID}$ simulates a practical, multi-authority timestamp system. This reflects real-world scenarios where multiple authorities and verifiers might provide timestamp services, necessitating verification of each party's organisation information and legitimacy to prevent malicious actors from generating fraudulent timestamps.

\textbf{Step 3. Timestamp verification.}

(a) TSA generates a random bit string $str_2$ with length $m$ using local QNRG, and an irreducible polynomial $q(x)$ using $str_2$. The hash function $h_{q, u}$ is obtained by $u=u_1 \oplus u_2$ and $q(x)$. $\operatorname{TS}(doc)$ is hashed by $h_{q,u}$ to get the hash value $h_{q, u} \left ( \operatorname{TS}(doc) \right )$. TSA encrypts $h_{q, u} \left ( \operatorname{TS}(doc) \right )$ and $str_2$ with the key $v=v_1 \oplus v_2$ and sends the cypher-text $v \oplus \{ h_{q, u} \left ( \operatorname{TS}(doc) \right )~|~str_2 \}$ and the timestamp document $\operatorname{TS}(doc)$ to the subscriber. Then the subscriber forwards the ciphertext and the timestamp document to the verifier.

(b) After they both receive the ciphertext and the timestamp document, the subscriber and the verifier exchange their keys $u_1, u_2, v_1, v_2$ through authenticated classical channels. Each of them can combine $u_1$ and $u_2$ to get $u^{\prime}, v_1$ and $v_2$ to get $v^{\prime}$. The subscriber and the verifier then decrypt the ciphertext by $v^{\prime}$ to obtain $h_{q, u} \left ( \operatorname{TS}(doc) \right )$ and $str_2$. The polynomial $q^{\prime}(x)$ is generated using $str_2$ and the hash function $h_{q, u}^{\prime}$ is obtained using $q^{\prime}(x)$ and $u^{\prime}$. The subscriber and the verifier apply $h_{q, u}^{\prime}$ to $\operatorname{TS}(doc)$ and verify the equivalence of the hash value $h_{q, u}^{\prime}\left (\operatorname{TS}(doc)\right )$ with $h_{q, u} \left ( \operatorname{TS}(doc) \right )$. If they are the same, the authentication succeeds. If they are different, the authentication fails.

(c) The timestamp is accepted only when both the subscriber and the verifier successfully verify it.

\textcolor{black}{Note that the communication between the verifier and CA during file verification, and between the subscriber and verifier during timestamp verification, should utilize classical authenticated  channels.}

\section{Security analysis}\label{sec:3}
The security of our protocol rests on the secrecy \textcolor{black}{and asymmetric relationship of correlated quantum keys, with the resulting non-repudiation and unforgeability properties of the file and timestamp.} We analyze the security of each step of the scheme, considering both the internal dishonest party and the external adversary. As \textcolor{black}{at most} three parties are involved \textcolor{black}{in each verification step}, only one internal dishonest party can be allowed because two dishonest parties can perfectly collude to defraud the third party~\cite{Weng2021secure,weng2023beating}. \textcolor{black}{We present the security analysis with perfect keys here. Our protocol can also employ imperfect keys obtained from a QKG without a privacy amplification step. The condition with imperfect keys is almost the same, except for minor differences in the specific security parameter for unforgeability. We note that in the protocol without privacy amplification, the same security as that using privacy amplification can be achieved by simply increasing the number of bits consumed, so security is not affected~\cite{li2023one}.} 

\subsection{Security of file verification}

\textcolor{black}{ File verification involves a three-party certification process using the messaging stage of the OTUH-QDS~\cite{yin2023experimental,li2023one}, which must account for the possibility of denial by the subscriber and tampering by the verifier. } 

\textcolor{black}{A dishonest subscriber may deny the $doc$ that both the verifier and CA have accepted, and claim that the timestamp created in the next stage is for another file. However, successful denial requires errors during the key exchange between the verifier and CA, as both the verifier and CA with correlated quantum keys must independently compute identical hash values of the file to validate it. The probability of successful denial is 
\begin{equation}   \varepsilon^{\operatorname{den}}_{\operatorname{FV}} = 2\varepsilon_{\operatorname{CC}},
\end{equation}
where $\varepsilon_{\operatorname{CC}}$ represents the failure probability of the classical communication, which is negligible (typically $10^{-11}$) and usually ignored in comprehensive communication systems~\cite{yin2023experimental}. Therefore, the non-repudiation probability makes our protocol secure under the denial attack from a dishonest subscriber.}

\textcolor{black}{On the other hand, a dishonest verifier} aims to intercept the transmitted data via classical channel and alter the transmitted information, $str_1$, the hash value $h_{p,s}(doc)$, and the file $doc$, into a new, self-consistent set: $str_1^{\prime}$, $h_{p^{\prime},s}(doc^{\prime})$, and $doc^{\prime}$. This is prevented by the unforgeability property, and Ref.~\cite{yin2023experimental} has proven that an optimal attack strategy requires knowledge of $str_1$. This necessitates guessing half of the key of $r$ (with length $n$), used to encrypt $str_1$. The probability of successful guessing is $P_{guess} = 2^{-n}$ for perfectly secret keys. The probability of a successful \textcolor{black}{tampering attack on file verification is
\begin{equation}
\varepsilon_{\operatorname{FV}}^{\operatorname{tam}} = L\cdot 2^{1-n}.
\end{equation}
}

Therefore, the probability of successful timestamp verification, $P_{FV}$, is given by
\begin{equation}
    P_{\operatorname{FV}} = \left(1-\varepsilon_{\operatorname{FV}}^{\operatorname{den}}\right )\left(1-\varepsilon_{\operatorname{FV}}^{\operatorname{tam}}\right )\ge 1-\left(\varepsilon_{\operatorname{FV}}^{\operatorname{den}}+\varepsilon_{\operatorname{FV}}^{\operatorname{tam}}\right ).
\end{equation}
where the scaling is tight enough because both $\varepsilon_{\operatorname{FV}}^{\operatorname{den}}$ and $\varepsilon_{\operatorname{FV}}^{\operatorname{tam}}$ are exceedingly small. The overall failure probability for the file verification step, $\varepsilon_{\operatorname{FV}}$, can therefore be approximated as the sum of the individual failure probabilities: $\varepsilon_{\operatorname{FV}} = \varepsilon_{\operatorname{FV}}^{\operatorname{den}} + \varepsilon_{\operatorname{FV}}^{\operatorname{tam}}$.

\subsection{Security of timestamp verification}

Timestamp verification involves a three-party certification process, which needs to consider the denial by TSA and tampering by any of the other two participants (the subscriber or the verifier) and an external adversary.

A dishonest TSA might falsely deny having issued a valid timestamp, even if both the subscriber and verifier have independently verified its authenticity. \textcolor{black}{Similar to file verification,} successful denial requires errors during the key exchange between the subscriber and the verifier, \textcolor{black}{whose probability is 
\begin{equation}
\varepsilon^{\operatorname{den}}_{\operatorname{TV}} = 2\varepsilon_{\operatorname{CC}}.
\end{equation}
}

A dishonest subscriber, verifier or external adversary may try to tamper with the transmitted information to cheat. The scenario is also the same as the analysis of the security of file verification, which is prevented by the unforgeability property. Since the attacker also needs to know $str_2$ for the implementation of optimal cheating methods, we can denote the probability of tampering as\textcolor{black}{
\begin{equation}
\varepsilon^{\operatorname{tam}}_{\operatorname{TV}} = k\cdot 2^{1-m}.
\end{equation}}

Successful timestamp verification demands that both malicious denial and tampering fail. The probability of successful file verification is
\begin{equation}
    P_{\operatorname{TV}} = \left(1-\varepsilon^{\operatorname{den}}_{\operatorname{TV}}\right )\left(1-\varepsilon^{\operatorname{tam}}_{\operatorname{TV}}\right )\ge 1-\left(\varepsilon^{\operatorname{den}}_{\operatorname{TV}}+\varepsilon^{\operatorname{tam}}_{\operatorname{TV}}\right ),
\end{equation}
and the overall failure probability for the timestamp verification is $\varepsilon_{\operatorname{TV}} = \varepsilon^{\operatorname{den}}_{\operatorname{TV}}+\varepsilon^{\operatorname{tam}}_{\operatorname{TV}}$.

\subsection{Composable security}
Our quantum timestamp protocol's security depends not only on the success of the file and timestamp verification steps but also on the underlying QKG stage. All components must function correctly to guarantee secure timestamps. The QKG stage involves estimating parameters such as bit and phase error rates, subject to statistical fluctuations~\cite{lucamarini2018overcoming,wang2018twin-field,xie2022breaking,Li2024finite}, which are used for subsequent error correction and privacy amplification. Although our scheme can employ imperfect keys without privacy amplification, the error correction step is still needed, and these parameters should also be estimated. The security bound of the QKG stage, denoted as $\varepsilon_{\operatorname{QKG}}$, must account for both perfect and imperfect key scenarios. Therefore, the overall security bound $\varepsilon_{\operatorname{sec}}$ for our quantum timestamp scheme considering QKG stage~\cite{korzh2015provably} is
\begin{equation}
    \varepsilon_{\operatorname{sec}} = \varepsilon_{\operatorname{QKG}} + \varepsilon_{\operatorname{FV}} + \varepsilon_{\operatorname{TV}}.
\end{equation}

\begin{figure}[t]
    \centering
    \includegraphics[width=82mm]{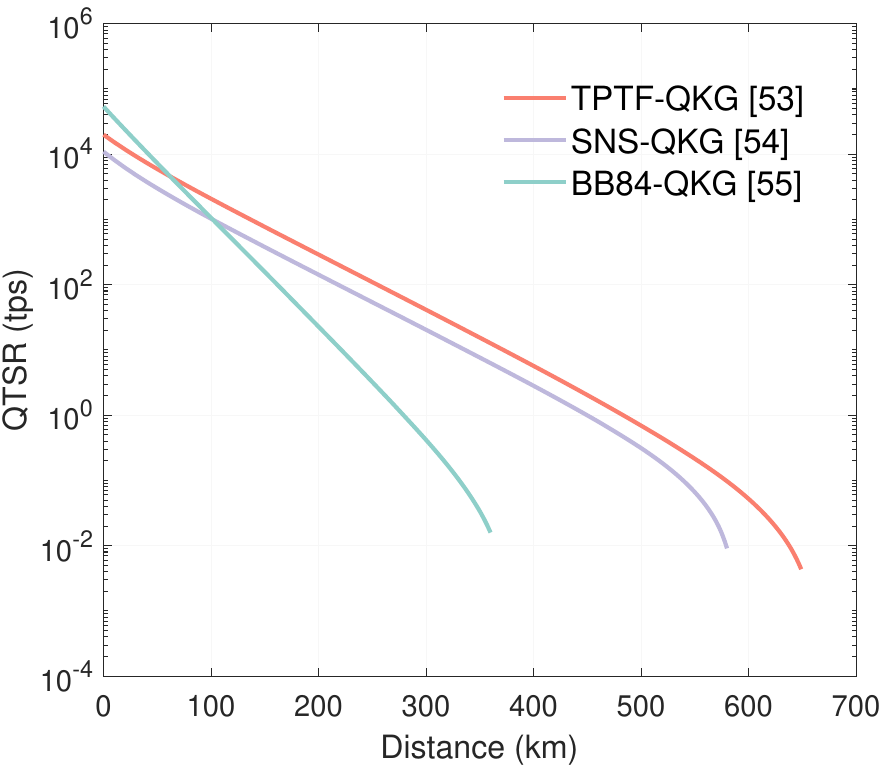}

    \caption{The efficiency of the proposed timestamp scheme. We use QTSR to denote the number of timestamps generated per second. The results of TPTF-QKG~\cite{xie2023Scalable}, SNS-QKG~\cite{jiang2019unconditional} and BB84-QKG~\cite{lim2014concise} are presented. Our protocol enables more than $10^2$-$10^3$ times of timestamping per second at intercity distances (100-300 km), and is available over 600 km using TPTF-QKG.}
    \label{fig2}
\end{figure}

\section{Numerical simulation}\label{sec:4}
This section presents numerical simulations of the proposed quantum timestamp scheme, providing a quantitative assessment of its efficiency. A security threshold is established by setting  $\varepsilon_{\operatorname{QKG}} = \varepsilon_{\operatorname{FV}} = \varepsilon_{\operatorname{TV}} = 10^{-10}$. The simulations incorporate realistic experimental parameters: detector efficiency $\eta_{de} = 70\%$, dark count rate $p_d = 10^{-8}$, misalignment error rate $e_t = 2\%$, and error correction efficiency $f_{\operatorname{cor}} = 1.1$. For simplicity, the channel loss and distance between each pair of the four parties are assumed to be equal. To reflect practical scenarios, the simulations model the timestamp of a 100 Mb file ($doc$) with a 1 Kb timestamp. The total number of pulses is set to $10^{14}$ to reflect the finite-key condition. We quantify the efficiency of a timestamp scheme by the quantum timestamp rate (QTSR), defined as the number of secure quantum timestamps generated per second. This can be calculated by\textcolor{black}{
\begin{equation}
    \text{QTSR} =\frac{\min (\text{KGR})}{l_{\text{key}}} = \frac{\min (\text{KGR})}{6(n+m)},
\end{equation}}
where KGR stands for the key generation rate and \textcolor{black}{$l_{\text{key}}=6(n+m)$} is the length of quantum keys needed to complete a timestamp stage, \textcolor{black}{including two $3n$-bit ($s_1$, $r_1$ and $s_2$, $r_2$) and two $3m$-bit ($u_1$, $v_1$ and $u_2$, $v_2$) quantum keys}. While the distances between each party are set to be the same, to account for channel variations among parties-resulting in differing channel losses and KGR, a conservative lower bound on QTSR is established by using the minimum KGR observed across all channels.

Assuming a laser repetition rate of 1 GHz and a fiber-loss coefficient of $\alpha = 0.165 \, \text{dB/km}$, Figure~\ref{fig2} depicts the calculated QTSR for transmission distances ranging from 0 to over 600 km (where key generation is feasible). The results demonstrate a QTSR exceeding 100 timestamps per second for inter-city distances (approximately 100 to 300 km). Specifically, when employing TPTF-QKG, $10^4$ times of successful stamping can be achieved, and our protocol is still available over 600 km spatial distance. Our simulations show the practicality and high efficiency of our protocol, building a theoretical foundation for the experimental implementation of quantum timestamps.

\section{Discussion and conclusion}\label{sec:5}
Secure timestamp protocols are essential for applications demanding verifiable chronological ordering of events. Classical schemes, however, are vulnerable to quantum attacks. This work proposes a quantum timestamp protocol offering information-theoretic security against such threats. The protocol employs one-time universal$_2$ hashing combined with one-time pad encryption using quantum keys, and can also time-stamp a document of arbitrary length with high efficiency. Our simulations, which quantify the timestamp generation rate, highlight the efficiency of our work, particularly over intercity distances. The practicality of our protocol is further enhanced by its simple experimental requirements. It utilizes only coherent quantum states, avoiding the complexities of high-dimensional qudits or entangled states, and is compatible with various QKG methods. 

\textcolor{black}{Leveraging these advantages, our information-theoretically secure quantum timestamping protocol represents a significant step forward for applications that require high security and integrity of time records against quantum attacks, such as digital financial systems, distributed databases, and military command infrastructures. Its practical design enables seamless integration into existing quantum networks, meeting the demands for secure timestamps in the quantum era.}

\begin{acknowledgments}
This work was supported by the National Natural Science Foundation of China (No. 12274223), the Program for Innovative Talents, Entrepreneurs in Jiangsu (No. JSSCRC2021484) and the Key Research and Development Program of Nanjing Jiangbei New Area (No. ZDYD20210101). We gratefully acknowledged the free icons designed by Flaticon from \href{https://www.flaticon.com/}{www.flaticon.com}.
\end{acknowledgments}

\newpage
\appendix

\section{\label{Sec:A1} One-time universal$_2$ hash function}

A family of hash functions, $\mathcal{H}$, mapping a set $\mathbb{P}$ to a set $\mathbb{Q}$, is defined as universal$_2$~\cite{carter1979universal} if, for any distinct $\mathcal{M}, \mathcal{M}^{\prime} \in \mathbb{P}$:
\begin{equation}
    \text{Pr}_{h \in \mathcal{H}}[h(\mathcal{M}) = h(\mathcal{M}^{\prime})] \le \frac{1}{|\mathbb{Q}|}
\end{equation}
where $|\mathbb{Q}|$ denoting the number of elements in set $\mathbb{Q}$. Universal$_2$ hash functions efficiently map arbitrarily long inputs to short hash values with low collision probability. Traditional random matrices form a universal$_2$ family, requiring $i \cdot j$ random bits to map a $j$-bit input to an $i$-bit hash. To reduce this cost, Toeplitz matrices~\cite{menezes2018handbook} are used instead, requiring only $i + j - 1$ random bits and exhibiting a collision probability of $2^{-i}$. Moreover, the LFSR-based Toeplitz matrices~\cite{krawczyk1994lfsr} offer another alternative, acting as an almost universal$_2$ hash function. These matrices are defined by an irreducible polynomial of degree $i$ over GF(2) and an $i$-bit random initial vector, and their collision probability is $j/2^{i-1}$. 

Analogous to one-time pad encryption, where a shared key is used only once to encrypt a message, the one-time universal$_2$ hash function proposed in Ref.~\cite{yin2023experimental} employs a randomly chosen initial vector and irreducible polynomial for each hash computation. This approach enhances security by making each hash computationally infeasible to predict or reverse engineer.  Participating parties share the keys used to generate the universal$_2$ hash function, with these keys discarded after a single use within the protocol.

\section{\label{Sec:A2} Generation of an irreducible polynomial}

The generation of irreducible polynomials is a critical step in various cryptographic applications, particularly those utilizing LFSR-based Toeplitz matrices for authentication.  An irreducible polynomial of degree $i$ over the GF(2) is a polynomial that cannot be factored into polynomials of lower degree over the same field. Such polynomials are represented in the standard form: $p(x) = x^i + a_{i-1}x^{i-1} + \cdots + a_1x + a_0$, where each coefficient is binary. This polynomial can be concisely represented by an $i$-bit string, $str = (a_{i-1}, a_{i-2}, \dots, a_1, a_0)$, where each bit corresponds to a coefficient. Since $a_0$ is always 1 for irreducible polynomials, only $i-1$ random bits are actually needed, effectively using the string $str = (a_{i-1}, \dots, a_1, 1)$. A QRNG is employed to efficiently generate these $i-1$ random bits, forming the basis for constructing candidate irreducible polynomials. Subsequently, an irreducibility test is performed on each generated candidate polynomial. A polynomial $p(x)$ of degree $i$ over GF(2) is irreducible if and only if it satisfies the following two conditions:

\begin{equation}
\begin{cases}
\text{(A)} \quad x^{2^i} \equiv x \pmod{p(x)} \\
\text{(B)} \quad \gcd(x^{2^{i/d}} - x, p(x)) = 1
\end{cases}
\end{equation}
where $d$ is any prime factor of $i$ and $\gcd(a(x), b(x))$ means the greatest common divisor (GCD) of two polynomials $a(x)$ and $b(x)$.  The efficient computation of both conditions (A) and (B) leverages optimized algorithms such as fast modular exponentiation (for condition A) and the extended Euclidean algorithm (for condition B)~\cite{von2013Modern}. If a generated polynomial does not satisfy both conditions, a new polynomial is generated and tested until an irreducible polynomial is found.

However, this method may be computationally expensive and require many random numbers. A more efficient method for generating irreducible polynomials, proposed in Ref.~\cite{Shoup1996OnFast}, can be adopted, assuming participants have a pre-stored irreducible polynomial $p_1(x)$ of order $i$. This method generates a new irreducible polynomial using only one $i$-bit random string.  This string is used as the coefficients of a random polynomial $p_2(x)$.  Then, $2i$ values are computed: $e_0 = f_0(0)$, $e_1 = f_1(0)$, ..., $e_{2n-1} = f_{2i-1}(0)$, where $f_j(x) = p_2^j(x) \pmod{p_1(x)}$ and $j \in \{0, 1, \dots, 2i-1\}$. These values can be efficiently computed using the Berlekamp-Massey algorithm~\cite{Massey1999Shift} to determine the $i$-order minimal polynomial of $p_2(x)$, which is irreducible.

\section{\label{Sec:A3} Construction of the LFSR-Based Toeplitz Hash matrix}

The LFSR-based Toeplitz hash function, a crucial component of our authentication mechanism, leverages the properties of LFSR-based Toeplitz matrices to efficiently map a $j$-bit message, $M = (M_0, M_1, \dots, M_{j-1})^T$ (the superscript means the transpose of the vector), to an $i$-bit hash value. The $i \times j$ Toeplitz matrix, $H_{i \times j}$, is dynamically generated based on the $i$-bit coefficient string $str$ of an irreducible polynomial $p(x)$, and an $i$-bit binary string $K = (k_{i-1},\cdots,k_1,k_0)$ obtained by QKG. The function operates as
\begin{equation}
    h_{p,K}(M) = H_{i \times j} \cdot M  \pmod 2,
\end{equation}
where $\cdot$ denotes the matrix multiplication and $\pmod 2$ means every element of the result is calculated modulo 2.

Now we describe the procedure to construct such a hash function. The LFSR is initialized to the state vector $S_1$, which is set to be $(k_{i-1},\cdots,k_1,k_0)^{T}$. Then, the LFSR generates a sequence of $i$-bit vectors with $S_1$ being the first by shifting down every element of the previous vector and adding a new element at the top. More specifically, the LFSR generates the following vectors:
\begin{equation}
\begin{aligned}
    &S_2 = (str\cdot S_1, k_{i-1},\cdots,k_1  )^T\pmod 2,\\
    &S_3 = (str\cdot S_2,str\cdot S_1, k_{i-1},\cdots,k_2  )^T\pmod 2,\\
    &\cdots,\\
    &S_j = (str\cdot S_{j-1},str\cdot S_{j-2},\cdots,str\cdot S_{j-i})^T\pmod 2.
\end{aligned}
\end{equation}
Finally, the LFSR-based Toeplitz hash matrix is constructed as
\begin{equation}
    H_{i\times j} = (S_1,S_2,\cdots,S_j).
\end{equation}
The use of an irreducible polynomial and the inherent randomness from the QRNG ensure the almost universal$_2$ property of this hash function. The specific collision probability is $j\cdot 2^{1-i}$~\cite{krawczyk1994lfsr}. This construction ensures efficient computation while maintaining a low collision probability, making it suitable for cryptographic applications.

\section{Simulation details\label{Sec:A4}}
The following sections detail the numerical simulation calculations for TPTF-QKG, BB84-QKG, and SNS-QKG, both with and without privacy amplification. For clarity, the calculations are presented for the generation of a single $n$-bit string. The simulation of generating $m$-bit string is simply to substitute the corresponding parameters.

\subsection{TPTF-QKG}
The TPTF-QKG calculation follows the TPTF-QKD method of Ref.~\cite {xie2023Scalable}. For intensities ($k_1$, $k_2$) and phase difference $\theta$, the left ($L$) and right ($R$) single-detector click gains are:

\begin{equation}
\begin{aligned}
Q_{k_1 k_2}^{L\theta} &= y_{k_1 k_2} \left( e^{\omega_{k_1 k_2} \cos\theta} - y_{k_1 k_2} \right), \\
Q_{k_1 k_2}^{R\theta} &= y_{k_1 k_2} \left( e^{-\omega_{k_1 k_2} \cos\theta} - y_{k_1 k_2} \right),
\end{aligned}
\end{equation}
where $y_{k_1 k_2} = e^{-(\eta_1 k_1 + \eta_2 k_2)/2} (1 - p_d)$, $\omega_{k_1 k_2} = \sqrt{\eta_1 k_1 \eta_2 k_2}$, $\eta_i (i\in \{1,2\})$ are the channel transmittances for $U_1$ and $U_2$, and $p_d$ is the dark count probability. The overall gain is

\begin{equation}
\begin{aligned}
    Q_{k_1 k_2} &= \frac{1}{2\pi} \int_0^{2\pi} (Q_{k_1 k_2}^{L\theta} + Q_{k_1 k_2}^{R\theta}) d\theta \\
    &= 2y_{k_1 k_2} \left[ I_0(\omega_{k_1 k_2}) - y_{k_1 k_2} \right],
\end{aligned}
\end{equation}
where $I_0(x)$ is the zeroth-order modified Bessel function of the first kind. The total event count for $\{k_1, k_2\}$ is $x_{k_1 k_2} = N_{\text{tot}} p_{k_1} p_{k_2} Q_{k_1 k_2}$.

We define the correct $Z$-basis events as $\{\mu_1 \mathbf{o}_1, \mathbf{o}_2 \mu_2\}$, $\{\mathbf{o}_1 \mu_1, \mu_2 \mathbf{o}_2\}$, and incorrect $Z$-basis events as $\{\mu_1 \mathbf{o}_1, \mu_2 \mathbf{o}_2\}$, $\{\mathbf{o}_1 \mu_1, \mathbf{o}_2 \mu_2\}$. The Z-basis correct ($n_C^Z$) and incorrect ($n_E^Z$) event counts are calculated as

\begin{align}
n_C^Z &= x_{\text{min}} \frac{x_{\mathbf{o}_1 \mu_2}}{x_0} \frac{x_{\mu_1 \mathbf{o}_2}}{x_1} = \frac{x_{\mathbf{o}_1 \mu_2} x_{\mu_1 \mathbf{o}_2}}{x_{\text{max}}}, \\
n_E^Z &= x_{\text{min}} \frac{x_{\mathbf{o}_1 \mathbf{o}_2}}{x_0} \frac{x_{\mu_1 \mu_2}}{x_1} = \frac{x_{\mathbf{o}_1 \mathbf{o}_2} x_{\mu_1 \mu_2}}{x_{\text{max}}},
\end{align}
where $x_0 = x_{\mathbf{o}_1 \mu_2} + x_{\mathbf{o}_1 \mathbf{o}_2}$, $x_1 = x_{\mu_1 \mathbf{o}_2} + x_{\mu_1 \mu_2}$, $x_{\text{min}} = \min(x_0, x_1)$, $x_{\text{max}} = \max(x_0, x_1)$.  $s_{11}^Z$ means the count of single-photon clicks for the $Z$-basis. The Z-basis bit error rate is $E^Z = m^Z / n^Z$, where $m^Z = (1 - e_d^Z) n_E^Z + e_d^Z n_C^Z$ denoting the number of $Z$-basis bit errors, and $n^Z = n_C^Z + n_E^Z$ is the total number of $Z$-basis events and $e_d^Z$ is the misalignment error.

The effective X-basis event count ($n^X$) and error count ($m^X$) are:

\begin{equation}
	\begin{aligned}
		n^X=&\frac{1}{\pi}\int_{0}^{\delta} x_{\nu_1\nu_2}^{\theta}d\theta =\frac{N_{\rm{tot}}p_{\nu_{a}}p_{\nu_{b}}}{\pi}\int_\sigma^{\sigma+\delta} y_{\nu_1\nu_2}\\
  & \times (e^{\omega_{\nu_1\nu_2}\cos\theta} +e^{-\omega_{\nu_1\nu_2}\cos\theta}-2y_{\nu_1\nu_2})d\theta.
	\end{aligned}
\end{equation}
and

\begin{equation}
	\begin{aligned}
		m^{X}&=\frac{1}{\pi}\int_\sigma^{\sigma+\delta} x_{\nu_1\nu_2}^\theta p_{E}d\theta =\frac{2N_{\rm{tot}}p_{\nu_{a}}p_{\nu_{b}}}{\pi}\int_\sigma^{\sigma+\delta} y_{\nu_1\nu_2}\\
		&\times\left[\frac{(1-y_{\nu_1\nu_2})^{2}}{e^{\omega_{\nu_1\nu_2}\cos\theta}+e^{-\omega_{\nu_1\nu_2}\cos\theta}-2y_{\nu_1\nu_2}}-1\right]d\theta,\\
	\end{aligned}
\end{equation}
where $p_E = \frac{2q_{\nu_1 \nu_2}^{L\theta} q_{\nu_1 \nu_2}^{R\theta}}{q_{\nu_1 \nu_2}^\theta q_{\nu_1 \nu_2}^\theta}$.

$x^*$, $\underline{x}$, and $\overline{x}$ denote the expected value, lower bound, and upper bound of $x$, respectively.  $x_{k_1k_2}$ represents the number of $\{k_1, k_2\}$ events. For post-matched events $\{k_1^{l_1}k_1^{l_2}, k_2^{l_1}k_2^{l_2}\}$, $n_{k_1^{l_1}k_1^{l_2}, k_2^{l_1}k_2^{l_2}}$ and $m_{k_1^{l_1}k_1^{l_2}, k_2^{l_1}k_2^{l_2}}$ denote the event count and error count, respectively.  For brevity, when $k_1^{l_1} = k_1^{l_2}$ and $k_2^{l_1} = k_2^{l_2}$, we use $2k_1, 2k_2$.

First, we calculate the following parameters:

1. $\underline{s}_{11}^Z$: Lower bound on the number of single-photon pairs in the Z-basis. This is calculated from the lower bounds of the expected values of $z_{10}$ ($z_{01}$), which denotes the times where $U_1$ ($U_2$) emits a single photon and $U_2$ ($U_1$) emits a vacuum state in an $\{\mu_1, \mathbf{o}_2\}$ ($\{\mathbf{o}_1, \mu_2\}$) event:

\begin{align}
    \underline{z}_{10}^* = N_{\text{tot}} p_{\mu_1} p_{\mathbf{o}_2} \mu_1 e^{-\mu_1} \underline{y}_{10}^*,\\
    \underline{z}_{01}^* = N_{\text{tot}} p_{\mathbf{o}_1} p_{\mu_2} \mu_2 e^{-\mu_2} \underline{y}_{01}^*,
\end{align}
where $\underline{y}_{10}^*$ and $\underline{y}_{01}^*$ are the yields, estimated using the decoy-state method:

\begin{equation}
\begin{aligned}
    \underline{y}_{01}^*\geq & \frac{\mu_2}{N_{\rm{tot}}(\mu_2\nu_2-\nu_2^2)} \\
    &\left(\frac{e^{\nu_2}\underline{x}_{o_1\nu_2}^{*}}{p_{o_1}p_{\nu_2}}\nonumber-\frac{\nu_2^2}{\mu_2^2}  \frac{e^{\mu_2}\overline{x}_{\hat{\mathbf{o}}_{a}\mu_2}^{*}}{p_{\hat{\mathbf{o}}_{a}}p_{\mu_2}} - \frac{\mu_2^2-\nu_2^2}{\mu_2^2}\frac{\overline{x}_{o o}^{d*}}{p_{o_1o_2}^d}\right),
\end{aligned}
\end{equation}

\begin{equation}
\begin{aligned}
    \underline{y}_{10}^*\geq & \frac{\mu_1}{N_{\rm{tot}}(\mu_1\nu_1-\nu_1^2)} \\
    &\left( \frac{e^{\nu_1}\underline{x}_{\nu_1o_2}^{*}}{p_{\nu_1}p_{o_{b}}}\nonumber-\frac{\nu_1^2}{\mu_1^2} \frac{e^{\mu_1}\overline{x}_{\mu_1\hat{\mathbf{o}}_{b}}^{*}}{p_{\mu_  a}p_{\hat{\mathbf{o}}_{b}}}- \frac{\mu_1^2-\nu_1^2}{\mu_1^2}\frac{\overline{x}_{o o}^{d*}}{p_{o_1o_2}^d}\right),\label{eq_decoy_Y01}
\end{aligned}
\end{equation}
where $x_{\mathbf{o}\mathbf{o}}^d = x_{\hat{\mathbf{o}}_1 \hat{\mathbf{o}}_2} + x_{\hat{\mathbf{o}}_1 \mathbf{o}_2} + x_{\mathbf{o}_1 \hat{\mathbf{o}}_2}$ is the count of events with at least one declared-vacuum state is sent, and $p_{\mathbf{o}\mathbf{o}}^d = p_{\hat{\mathbf{o}}_1} p_{\hat{\mathbf{o}}_2} + p_{\hat{\mathbf{o}}_1} p_{\mathbf{o}_2} + p_{\mathbf{o}_1} p_{\hat{\mathbf{o}}_2}$ is the its probability. The lower bound of $s_{11}^{Z*}$ is

\begin{equation}
\underline{s}_{11}^{Z*} = \frac{\underline{z}_{10}^* \underline{z}_{01}^*}{x_{\text{max}}}.
\end{equation}

2. $\underline{s}_{0\mu_2}^Z$: The number of $Z$-basis events where $U_1$ sends a vacuum state in both time bins, and $U_2$'s total intensity is $\mu_2$. Let $z_{00}$ ($z_{0\mu_2}$) be the number of events where $U_1$'s state measured to vacuum state in an $\{\mu_1, \mathbf{o}_2\}$ ($\{\mu_1, \mu_2\}$) event. The lower bounds on the expected values are: $\underline{z}_{00}^*={ p_{\mu_1} p_{\mathbf{o}_2}e^{-\mu_1}\underline{x}_{o o}^{d*}}/{p_{o_1o_2}^d}$ and $ ~\underline{z}_{0\mu_2}^*=  {p_{\mu_1} p_{\mu_2}e^{-\mu_1}\underline{x}_{\mathbf{o}_{a}\mu_2}^{*}}/{p_{\mathbf{o}_{a}}p_{\mu_2}}$. Using $\underline{x}_{\mathbf{o}_{a}\mu_2}^{*}={ p_{\mathbf{o}_{a}} \underline{x}_{\hat{\mathbf{o}}_{a}\mu_2}^{*}}/{ p_{\hat{\mathbf{o}}_{a}}}$, and $~\underline{x}_{\mathbf{o}_{a}\mathbf{o}_2}^{*}={ p_{\mathbf{o}_{a}} p_{\mathbf{o}_2}\underline{x}_{o o}^{d*}}/{p_{oo}^d}$, the lower bound on $s_{0\mu_2}^{Z*}$ is

\begin{equation}
\underline{s}_{0\mu_2}^{Z*} = \frac{\underline{x}_{\mathbf{o}_1 \mu_2}^* \underline{z}_{00}^*}{x_{\text{max}}} + \frac{\underline{x}_{\mathbf{o}_1 \mathbf{o}_2}^* \underline{z}_{0\mu_2}^*}{x_{\text{max}}}.
\end{equation}

3.  $\underline{s}_{11}^X$:  To determine the expected number of single-photon pairs in the X-basis, we consider the phase difference between $U_1$ and $U_2$, given by $\theta = \theta_1 - \theta_2 + \phi_{12}$. The X-basis events are categorized according to their phase difference $\theta$, where $\theta \in [-\delta, \delta] \cup [\pi - \delta, \pi + \delta]$, and $x_{k_1 k_2}^\theta$ represents the count of events with intensities $\{k_1, k_2\}$ and phase difference $\theta$.  Post-processing only retains pairs of events that share the same phase difference $\theta$.  Assuming a uniform distribution of $\theta$ and a misalignment angle $\sigma$, the expected number of single-photon pairs, $\underline{s}_{11}^X$, is calculated as follows.

\begin{equation}
	\begin{aligned}
		\underline{s}_{11}^{X*} &=\frac{1}{\pi}\int_\sigma^{\sigma+\delta} x_{\nu_1\nu_2}^\theta\times 2\frac{\nu_2e^{-(\nu_1+\nu_2)}\underline{y}_{01}^*}{q_{\nu_1\nu_2}^\theta}\frac{\nu_1e^{-(\nu_1+\nu_2)}\underline{y}_{10}^*}{q_{\nu_1\nu_2}^\theta}d \theta\\
		&=\frac{N_{\rm{tot}}p_{\nu_{a}}p_{\nu_{b}}}{\pi}\int_\sigma^{\sigma+\delta}\frac{2\nu_1\nu_2e^{-2(\nu_1+\nu_2)}\underline{y}_{01}^*\underline{y}_{10}^*}{q_{\nu_1\nu_2}^\theta},\\
	\end{aligned}
\end{equation}
This calculation uses the gain $q_{\nu_1 \nu_2}^\theta$, which represents the probability of a successful detection event (exactly one detector clicks) given that $U_1$ selects intensity $\nu_1$, $U_2$ selects intensity $\nu_2$, and the phase difference between their pulses is $\theta$. The total number of events with intensities $\{\nu_1, \nu_2\}$ and phase difference $\theta$ is given by $x_{\nu_1 \nu_2}^\theta = N_{\text{tot}} p_{\nu_1} p_{\nu_2} q_{\nu_1 \nu_2}^\theta$, where $N_{\text{tot}}$ is the total number of time bins, and $p_{\nu_1}$ and $p_{\nu_2}$ are the probabilities that $U_1$ and $U_2$ choose intensities $\nu_1$ and $\nu_2$, respectively.

4. $\overline{e}_{11}^{X}$:  The upper bound on the X-basis bit error rate is derived from the equivalence between the expected Z-basis phase error rate and the expected X-basis bit error rate for single-photon pairs. The upper bound on the number of single-photon pair errors in the X-basis, $\overline{t}_{11}^X$, is
\begin{equation}
	\begin{aligned}
		\overline{t}_{11}^X\leq& m^X - \underline{(m_{\nu_10,\nu_20}+m_{0\nu_1,0\nu_2})} +\overline{m}_{00,00}, 	
	\end{aligned}\label{eq_TPTFQKD _t11},
\end{equation}
where $m_{\nu_1 0, \nu_2 0}$ and $m_{0 \nu_1, 0 \nu_2}$ represent error counts for vacuum states in time bins $l_1$ and $l_2$ of $\{2\nu_1, 2\nu_2\}$ events, respectively, and $m_{00,00}$ is the error count for vacuum states in both time bins.  The expected values of these error counts are:
\begin{equation}
	\begin{aligned}	
		\underline{(n_{\nu_10,\nu_20}+n_{0\nu_1,0\nu_2})}^*=&\frac{2}{\pi}\int_\sigma^{\sigma+\delta} x_{\nu_1\nu_2}^\theta\frac{e^{-(\nu_1+\nu_2)}\underline{q}_{00}^{*}}{q_{ \nu_1\nu_2}^\theta}d\theta\\
		=&\frac{\delta N_{\rm{tot}}p_{\nu_{a}}p_{\nu_{b}}e^{-(\nu_1+\nu_2)}\underline{q}_{00}^{*}}{\pi},\\
	\end{aligned}
\end{equation}
and
\begin{equation}
	\begin{aligned}	
		\overline{n}_{00, 00}^*=&\frac{1}{\pi}\int_\sigma^{\sigma+\delta} x_{\nu_1\nu_2}^\theta\left(\frac{e^{-(\nu_1+\nu_2)}\overline{q}_{00}^{*}}{q_{ \nu_1\nu_2}^\theta}\right)^2d\theta\\
		=&\frac{N_{\rm{tot}}p_{\nu_{a}}p_{\nu_{b}}}{\pi}\int_\sigma^{\sigma+\delta}\frac{e^{-2(\nu_1+\nu_2)}(\overline{q}_{00}^{*})^2}{q_{\nu_1\nu_2}^\theta}d\theta,\\
	\end{aligned}
\end{equation}
respectively. Here $q_{00}^*=x_{o_1o_2}^{d*}/(N_{\rm{tot}}p_{oo}^d)$. Because the vacuum state error rate is 1/2, $\underline{(m_{\nu_1 0, \nu_2 0} + m_{0 \nu_1, 0 \nu_2})}^* = \frac{1}{2} \underline{(n_{\nu_1 0, \nu_2 0} + n_{0 \nu_1, 0 \nu_2})}^*$ and $\overline{m}_{00, 00}^* = \frac{1}{2} \overline{n}_{00, 00}^*$.  Therefore, we have:
\begin{equation}
\overline{e}_{11}^{X}=\overline{t_{11}^{X}}/\underline{s}_{11}^X.	
\end{equation}

5. $\overline{\phi}_{11}^{Z}$: The upper bound on the Z-basis phase error rate is obtained using random sampling without replacement~\cite{yin2020tight}, setting a failure probability $\varepsilon$:
\begin{equation}
	\begin{aligned}
		\overline{\phi}_{11}^{Z}\leq&\overline{e}_{11}^{X}+	\gamma^U \left(\underline{s}_{11}^Z,\underline{s}_{11}^X,\overline{e}_{11}^{X},\varepsilon\right),\\
	\end{aligned}\label{eq_TPTFQKD _phi11z}
\end{equation}
where
\begin{equation}
	\gamma^{U}(n_1,n_2,\lambda,\varepsilon)=\frac{\frac{(1-2\lambda)AG}{n_1+n_2}+
		\sqrt{\frac{A^2G^2}{(n_1+n_2)^2}+4\lambda(1-\lambda)G}}{2+2\frac{A^2G}{(n_1+n_2)^2}},
\end{equation}
with $A=\max\{n_1,n_2\}$ and $G=\frac{n_1+n_2}{n_1n_2}\ln{\frac{n_1+n_2}{2\pi n_1n_2\lambda(1-\lambda)\varepsilon^{2}}}$.

6. $\underline{s}_{11}^{Zn}$,~$\underline{s}_{0\mu_2}^{Zn}$ and~$\overline{\phi}_{11}^{Zn}$: \textcolor{black}{Within the $n$-bit string extracted from the Z-basis key, $\underline{s}_{0\mu_2}^{Zn}$ and $\underline{s}_{11}^{Zn}$ denote the lower bounds for the number of vacuum events and single-photon pair events, respectively. The upper bound for the phase error rate of these single-photon pairs is represented by $\overline{\phi}_{11}^{Zn}$. They are calculated by}
\begin{equation}
	\begin{aligned}
		\underline{s}_{11}^{Zn}\geq&n\left[\underline{s}_{11}^{Z}/n^Z-	\gamma^U \left(n,n^Z-n,\underline{s}_{11}^{Z}/n^Z,\epsilon\right)\right],\\
		\underline{s}_{0\mu_2}^{Zn}\geq&n\left[\underline{s}_{0\mu_2}^{Z}/n^Z-	\gamma^U \left(n,n^Z-n,\underline{s}_{0\mu_2}^{Z}/n^Z,\epsilon\right)\right],\\
		\overline{\phi}_{11}^{Zn}\leq&\overline{\phi}_{11}^{Z}+	\gamma^U \left(\underline{s}_{11}^{Zn},\underline{s}_{11}^{Z}-\underline{s}_{11}^{Zn},\overline{\phi}_{11}^{Z},\epsilon\right).\\
	\end{aligned}
\end{equation}

The length of the final key for TPTF-QKG with privacy amplification is
\begin{equation}\label{l}
	\begin{aligned}
		l_{\operatorname{key}} =& \underline{s}_{0\mu_2}^{Z}+\underline{s}_{11}^{Z} \left [1-H(\overline{\phi}_{11}^{Z})\right]-n^Z 
f_{\operatorname{cor}} H(E^Z) \\
		&-\log_{2}{\frac{2}{\epsilon_{cor}}}-2\log_{2}{\frac{1}{2\epsilon_{PA}}},
	\end{aligned}
\end{equation}
where $\epsilon_{PA}$ and $\epsilon_{cor}$ are the privacy amplification and error correction failure probability, respectively.

\subsection{SNS-QKG}

In the SNS-QKD protocol of Ref.~\cite{jiang2019unconditional}, Alice and Bob choose a signal window with probability $p_z$ and a decoy window with probability $1 - p_z$. They send a vacuum state with probability $p_{z0}$ or a phase-randomized weak coherent state of intensity $\mu_z$ with probability $1 - p_{z0}$ for the signal window. For the decoy window, they send a vacuum state with probability $p_0$, a coherent state of intensity $\mu_1$ with probability $p_1$, and a coherent state of intensity $\mu_2$ with probability $1 - p_0 - p_1$. Following the approach in Ref.~\cite {jiang2019unconditional}, $N_{jk}$ events are recorded for each intensity pair ($jk$), where $j$ and $k$ represent the intensities sent by $U_1$ and $U_2$, respectively, and $jk \in \{00, 01, 10, 02, 20\}$ (intensities 1 and 2 are used).  After sifting, $n_{jk}$ one-detector heralded events remain. The counting rate for each intensity pair is defined as $S_{jk} = n_{jk} / N_{jk}$.  Then:

\begin{equation}\label{eq28}
\begin{split}
N_{00}=&\left[(1-p_z)^2p_0^2+2(1-p_z)p_zp_0p_{z0}\right]N_{\rm{tot}},\\
N_{01}=&N_{10}=\left[(1-p_z)^2p_0p_1+(1-p_z)p_zp_{z0}p_1\right]N_{\rm{tot}},\\
N_{02}=&N_{20}=\big[(1-p_z)^2(1-p_0-p_1)p_0\\
&+(1-p_z)p_zp_{z0}(1-p_0-p_1)\big]N_{\rm{tot}}.
\end{split}
\end{equation}

To determine the upper bound of $e_1^{ph}$, two subsets of $X_1$ windows, denoted as $C_{\Delta^+}$ and $C_{\Delta^-}$, are defined. The number of instances within each subset, $N_{\Delta^\pm}$, is calculated as follows.

\begin{equation}
N_{\Delta^\pm} = \frac{\Delta}{2\pi} (1 - p_z)^2 p_1^2 N_{\rm{tot}}.
\end{equation}

Let $n_{\Delta^+}^R$ denote the number of effective events where the right detector clicks within the $C_{\Delta^+}$ subset, and $n_{\Delta^-}^L$ denote the number of effective events where the left detector clicks within the $C_{\Delta^-}$ subset. The overall counting error rate for these subsets, $T_\Delta$, is then calculated as
\begin{equation}
    T_\Delta = \frac{n_{\Delta^+}^R + n_{\Delta^-}^L}{2N_{\Delta^\pm}}.
\end{equation}
The lower bound for the expected counting rate of untagged single photons, denoted as  $\underline{s}_1^{Z*}$, is given by

\begin{equation}\label{s11l}
\begin{split}
s_1^{Z*}\ge &\underline{s}_1^{Z*}=\frac{1}{2\mu_1\mu_2(\mu_2-\mu_1)}\big[\mu_2^2e^{\mu_1}(\underline{S}_{01}^{*}+\underline{S}_{10}^{*})\\
&-\mu_1^2e^{\mu_2}(\overline{S}_{02}^{*}+\overline{S}_{20}^*)-2(\mu_2^2-\mu_1^2)\overline{S}_{00}^*\big].
\end{split}
\end{equation}
In this equation, $S_{jk}^*$ represents the expected value of the counting rate $S_{jk}$.  The upper and lower bounds of this expected value, estimated from the observed data, are denoted by $\overline{S}_{jk}^*$ and $\underline{S}_{jk}^*$, respectively. These bounds account for statistical uncertainties in the experimental measurements.

The upper bound for the expected phase-flip error rate of untagged single photons, denoted as $\overline{e}_1^{ph*}$, is expressed as

\begin{equation}
\overline{e}_1^{ph*} = \frac{\overline{T}_\Delta^* - \frac{1}{2} e^{-2\mu_1} \underline{S}_{00}^*}{2\mu_1 e^{-2\mu_1} \underline{s}_1^{Z*}},
\end{equation}
using the vacuum state error rate is 1/2.

With total transmittance $\eta$:

\begin{align}
&n_{00}=2p_d(1-p_d)N_{00},\\
&n_{01}=n_{10}=2\left[(1-p_d)e^{\eta\mu_1/2}-(1-p_d)^2e^{-\eta\mu_1}\right]N_{01},\\
&n_{02}=n_{20}=2\left[(1-p_d)e^{\eta\mu_2/2}-(1-p_d)^2e^{-\eta\mu_2}\right]N_{02},\\
&n^Z=n_{signal}+n_{error},\\
&E_z=\frac{n_{error}}{n_t},\\
&n_{\Delta^+}^R =n_{\Delta^-}^L=\left[T_X(1-2e_d)+e_dS_X\right]N_{\Delta^{\pm}},
\end{align}
where we have:
\begin{equation}
    \begin{aligned}
        n_{signal}=&4N_{\rm{tot}}p_z^2p_{z0}(1-p_{z0})\\
       &\big[(1-p_d)e^{-\eta\mu_z/2}-(1-p_d)^2e^{-2\eta\mu_z}\big],\\
        n_{error}=&2N_{\rm{tot}}p_z^2(1-p_{z0})^2\\
        &\big[(1-p_d)e^{-\eta\mu_z}I_0(\eta\mu_z)-(1-p_d)^2e^{-2\eta\mu_z}\big]\\
       &+2N_{\rm{tot}}p_z^2p_{z0}^2p_d(1-p_d),\\
    \end{aligned}
\end{equation}
and
\begin{equation}
    \begin{aligned}
        T_X&=\frac{1}{\Delta}\int_{-\frac{\Delta}{2}}^{\frac{\Delta}{2}}(1-p_d)e^{-2\eta\mu_1\cos^2{\frac{\delta}{2}}}d\delta-(1-p_d)^2e^{-2\eta\mu_1},\\
    S_X&=\frac{1}{\Delta}\int_{-\frac{\Delta}{2}}^{\frac{\Delta}{2}}(1-p_d)e^{-2\eta\mu_1\sin^2{\frac{\delta}{2}}}d\delta-(1-p_d)^2e^{-2\eta\mu_1}+T_X.
    \end{aligned}
\end{equation}

In SNS-QKG, the values of an $n$-bit key extracted from an $n^Z$-bit key are calculated as
\begin{equation}
	\begin{aligned}
		&\underline{s}_{1n}^{Z*}=n\left[\underline{s}_1^{Z*}/n^Z-	\gamma^U \left(n,n^Z-n,\underline{s}_1^{Z*}/n^Z,\epsilon\right)\right],\\
		&\overline{e}_{1n}^{ph*}= \overline{e}_1^{ph*}+	\gamma^U \left(\underline{s}_{1n}^{Z*},\underline{s}_1^{Z*}-\underline{s}_{1n}^{Z*} ,\overline{e}_1^{ph*},\epsilon\right) ,\\
	\end{aligned}
\end{equation}
and the amount of secret information is
\begin{equation}
    H_n=\underline{s}_{1n}^{Z*}\left[1-H(\overline{e}_{1n}^{ph*})\right] -\lambda_{EC},
\end{equation}
where $\lambda_{EC}=n f_{\operatorname{cor}} H(E_z)$. \textcolor{black}{The key length is then calculated by
\begin{equation}
\begin{aligned}
        l_{\operatorname{key}} =  &\underline{s}_{1}^{Z*}\left[1-H(\overline{e}_{1}^{ph*})\right] -\lambda_{EC}\\
        &-\log_2(\frac{2}{\varepsilon_{\operatorname{cor}}}) - 2\log_{2}{\frac{1}{\sqrt{2}\epsilon_{PA}\hat{\varepsilon}}}.
\end{aligned}
\end{equation}}

\subsection{BB84-QKG}
According to~\cite {lim2014concise}, the count of vacuum and single-photon clicks of the $X$-basis are
\begin{equation}\label{eqn2}
s_{X,0} \geq \tau_{0}\frac{\mu_\rd n_{X,\mu_\rdd}^--\mu_\rdd n_{X,\mu_\rd}^+}{\mu_\rd-\mu_\rdd},
\end{equation}
\begin{equation} \label{eqn3}
s_{X,1} \geq \frac{\tau_{1}\mu_\rs\left[n_{X,\mu_\rd}^--n_{X,\mu_\rdd}^+-\frac{\mu_\rd^2-\mu_\rdd^2}{\mu_\rs^2}(n_{X,\mu_\rs}^+- \frac{s_{X,0}}{\tau_0})\right]}{\mu_\rs(\mu_\rd-\mu_\rdd)-\mu_\rd^2+\mu_\rdd^2}.
\end{equation}
where $\tau_{n}:=\sum_{\mu\in\cK}e^{-\mu}\mu^n p_{\mu}/n!$ is the probability of the $n$-photon state in a pulse with intensity $\mu$, and
\[
n_{X,\mu}^\pm:=\frac{e^{\mu}}{p_{\mu}}\left(n_{X,\mu}\pm\sqrt{ \frac{n_X}{2}\log_2\frac{21}{\esec}}\right),~\forall~\mu \in \cK.
\]
We use $\mu_1$, $\mu_2$ and $\mu_3$ to represent the signal and two decoy intensities. 

Using Z-basis statistics and following a similar calculation procedure, the number of vacuum events ($s_{Z,0}$) and single-photon events ($s_{Z,1}$) for $\mathcal{Z} = \bigcup_{k \in \mathcal{K}} \mathcal{Z}_k$ are determined. The single-photon phase error rate in the X-basis, $\phi_{X,1}$, is then calculated as follows.

\begin{equation} \label{eqn5}
\phi_{X,1}:=\frac{c_{X,1}}{s_{X,1}}  \leq \frac{v_{Z,1}}{s_{Z,1}} + \gamma^{U}\left( s_{Z,1},{s}_{X,1},\frac{v_{Z,1}}{s_{Z,1}},\esec \right), 
\end{equation}
where
\[
v_{Z,1} \leq \tau_{1}\frac{m_{Z,\mu_\rd}^+-m_{Z,\mu_\rdd}^-}{\mu_\rd-\mu_\rdd},
\]
and
\[
m_{Z,k}^{\pm}:=\frac{e^{k}}{p_k}\left(m_{Z,k}\pm\sqrt{ \frac{m^Z}{2}\log_2\frac{21}{\esec}}\right),~\forall~k \in \cK.
\]

%\[
%\gamma\left(a,b,c,d \right):= \sqrt{\frac{(c+d)(1-b)b}{cd\log2}\log_2\left( \frac{c+d}{cd(1-b)b} \frac{21^2}{a^2}\right)}.
%\]

The total number of events in the X-basis is given by $n_X = \sum_{k \in \mathcal{K}} n_{X,k}$, and the total number of errors is $m_X = \sum_{k \in \mathcal{K}} m_{X,k}$. For a selected $n$-bit string in BB84-QKG, the following parameters are estimated by

\begin{align}
s_{X,0}^n &\geq n \left[ \frac{\underline{s}_{X,0}}{n^Z} - \gamma^U \left( n, n^Z - n, \frac{\underline{s}_{X,0}}{n^Z}, \epsilon \right) \right], \\
s_{X,1}^n &\geq n \left[ \frac{\underline{s}_{X,1}}{n^Z} - \gamma^U \left( n, n^Z - n, \frac{\underline{s}_{X,1}}{n^Z}, \epsilon \right) \right], \\
\phi_{X,1}^n &\leq \phi_{X,1} + \gamma^U \left( \underline{s}_{X,1}^n, \underline{s}_{X,1} - \underline{s}_{X,1}^n, \overline{\phi}_{X,1}, \epsilon \right).
\end{align}

Thus we have
\begin{equation}
    H_n = \underline{s}_{X,0}^n + \underline{s}_{X,1}^n \left[ 1 - H(\overline{\phi}_{X,1}^n) \right] - \lambda_{EC},
\end{equation}
where $\lambda_{EC} = n f_{\operatorname{cor}} H(m_X / n_X)$. \textcolor{black}{The key length is then calculated by
\begin{equation}
\begin{aligned}
    l_{\operatorname{key}} =& \underline{s}_{X,0} + \underline{s}_{X,1} \left[ 1 - H(\overline{\phi}_{X,1}^n) \right] - \lambda_{EC}\\
    &-\log_2(\frac{2}{\varepsilon_{\operatorname{cor}}}) - 6\log_{2}{\frac{22}{\epsilon_{PA}}}.
\end{aligned}
\end{equation}}

%\bibliography{TSref}


\end{document}